Title page:

# Title: Records of sunspot and aurora activity during 581–959 CE in Chinese official histories in the periods of *Suí*, *Táng*, and the Five Dynasties and Ten Kingdoms

Harufumi Tamazawa[1], Akito Davis Kawamura[1], Hisashi Hayakawa[2], Asuka Tsukamoto[3], Hiroaki Isobe[4,5], Yusuke Ebihara[4,6]

[1] Kwasan Observatory, Kyoto University, Kyoto, Japan

[2] Graduate School of Letters, Kyoto University

[3] Center for Collaborative Study with Community, Gifu University

[4] Unit of Synergetic Studies for Space, Kyoto University, Kyoto, Japan

[5] Graduate School of Advanced Integrated Studies in Human Survivability, Kyoto University, Kyoto, Japan

[6] Research Institute of Sustainable Humanosphere, Kyoto University

**Abstract**

Recent studies of radioisotopes in tree rings or ice cores suggest that extreme space weather events occurred in the pre-telescope age. Observational records of naked-eye sunspots and low-latitude auroras in historical documents in pre-telescopic age can provide useful information on past solar activity. In this paper, we present the results of a comprehensive survey of records of sunspots and auroras in Chinese official histories from the 6th century to the 10th century, in the period of *Suí*, *Táng*, the Five Dynasties and Ten Kingdoms.

These official histories contain records of continuous observations with well-formatted reports conducted under the policy of the government. A brief comparison of the frequency of observations of sunspots and auroras with the observations of radioisotopes as an indicator of solar activity during the corresponding periods is provided. Based on our data, we survey and compile the records of sunspots and auroras in historical documents from various locations and in several languages, and ultimately provide these as open data to the scientific community.

## 1. Introduction

Since the invention of the telescope in the early 17th century, the number of sunspots has been recorded continuously, providing an indicator of the variation of solar magnetic activity (Hathaway 2015). Solar magnetic activity that occurred in pre-telescopic age is of great interest to researchers of solar physics or space weather from the viewpoints of the physical origin of the solar magnetic field and its long-term variation, as well as the solar influence on the terrestrial climate, and it can be investigated by indirect proxies such as radioactive isotopes generated by cosmic rays (Usoskin 2013). An alternative source for information of past solar activity and its influence on Earth can be found in historical documents, where observations of naked-eye sunspots and low-latitude auroras by the sky-watchers in pre-telescopic age are recorded. The large plasma ejection from the Sun, coronal mass ejection (CME), with extraordinary large energy, cause aurora observed in low latitude area, where aurora is not observed usually, when CME hits the magnetosphere of the Earth (e.g. Pulkkinen 2007 for a review). These low latitude auroras are seen when solar activity is in an active phase, i.e. with large sunspots on the solar disc. Large sunspots have large magnetic energy and stored energy is released as solar flare, CME, and so on. Therefore, the records of observation of large (i.e. naked-eye) sunspots and low latitude aurora are probably good proxies of solar activity (Gonzalez et al. 1994; Shiokawa et al. 2005; Vaquero & Vázquez 2009; Usoskin 2013; Odenwald 2015).

Many authors have published lists of astronomical events, including naked-eye sunspots and low-latitude auroras, retrieved from historical documents from Japan (Kanda 1933; Matsushita 1956; Nakazawa et al. 2004; Shiokawa et al. 2005), Korea (Lee et al. 2004), Babylon (Stephenson et al. 2004; Hayakawa et al. 2016d), the West Asia (Basurah 2006; Hayakawa et al. 2016c), Western Europe (Fritz 1873; Link 1962; Dell'Dall'Olmo 1979; Stothers 1979; Vaquero & Trigo 2005; Vaquero et al. 2010), Russia (Vyssotsky 1949), North America (Broughton 2002), the Tropical Atlantic Ocean (Vázquez & Vaquero 2010), and China (Schove and Ho 1959; Keimatsu 1970–1976; Yau & Stephenson 1988; Saito & Ozawa 1992; Yau et al. 1995; Xu et al. 2000; Hayakawa et al. 2015; Kawamura et al. 2016; Hayakawa et al. 2016e, submitted). A review of the historical records of solar activity was given in the monograph by Vaquero & Vázquez (2009).

Interest in the historical records of sunspots and auroras has been boosted recently by the discovery of "superflares" on solar type stars (Schaefer et al. 2000; Maehara et al. 2012; Shibayama et al. 2013) and cosmic-ray events during the 8th and 10th centuries (Miyake et al. 2012; Miyake et al. 2013). The total energy of the superflares found in the Kepter data by Maehara et al. (2012) and Shibayama et al. (2013) ranges between $10^{33}$–$10^{35}$ ergs, which is 10–1000 times larger than that of the Oct 28, 2003 X17 flare, as reported in Schrijver et al. (2012). Notsu et al. (2015a; 2015b) reported that some of those superflare stars have large starspots and a relatively slow rotational velocity close to that of the sun, which suggests similar superflares may also occur in the present Sun (see Aulanier et al. 2013, Shibata et al. 2013 for a theoretical discussion on this possibility).

Miyake et al. (2012; 2013) discovered anomalous increases of atmospheric $^{14}$C in tree rings during 774–775 CE and during 993–994 CE. They strongly indicate the sharp increase in cosmic ray fluxes during those periods. These two events were confirmed by $^{14}$C measurements of several different tree rings (Usoskin et al. 2013; Jull et al. 2014; Güttler et al. 2013; Miyake et al. 2014), corals of the South China Sea (Liu et al. 2014), and the Antarctic Dome Fuji ice core (Miyake et al. 2015). Various suggestions for their origin have been put forward; an extreme solar proton event (SPE) (Usoskin and Kovaltsov 2012; Eichler and Mordecai 2012; Melott and Thomas 2012; Thomas et al. 2013; Cliver et al. 2014), a nearby supernova (Miyake et al. 2012), a gamma ray burst (Hambaryan & Neuhäuser 2013; Pavlov et al. 2013), and a cometary impact on Earth (Liu et al. 2014) have been considered. If one can identify clear evidence of the occurrence of extremely large sunspots or intense auroras in the historical records, it can provide a strong support for the "superflare" scenario of the $^{14}$C events. Several independent works have been recently published with regards to this scenario (Usoskin et al. 2013; Neuhäuser & Hambaryan 2014; Zhou et al.2014; Neuhäuser & Neuhäuser 2015; Chapman et al. 2015; Stephenson 2015; Hayakawa et al. 2016a; Hayakawa et al. 2016f).

The authors believe that those historical records should be used more widely as basic data for the scientific community. For this purpose, we have started a project in which provide a list of the observational records of sunspots and auroras in historical records, which are available online. The bibliographic information and the original text of corresponding passages are provided for the examination of reliability. Our previous paper (Hayakawa et al. 2015, hereafter referred to as H15) was our first step of this project, providing the results of the survey of the candidates of sunspot and aurora observations in *Sòngshǐ,* a Chinese official history covering the period during 10th–13th century (the *Sòng* dynasty era). We have extended our survey to 13th -20th century (Hayakawa et al. 2016e, submitted; Kawamura et al. 2016). In this paper, we extend our survey to the medieval Chinese official histories from 6th–10th century that cover the *Suí, Táng*, and the Five Dynasties and Ten Kingdoms eras. All the data provided in H15 and in this paper are available at http://www.kwasan.kyoto-u.ac.jp/~palaeo/.

## 2. Methods

### 2.1. Source Documents

The method used in this paper is the same as that of H15. We searched for the keywords usually associated with sunspots and auroras in imperial chronicles (本紀), in the Treatises of Astronomy (天文志), the Treatises of "Five Elements/Five Phases (五行志)[1]," and so on in Chinese official histories (正史). We then carefully inspected the corresponding section of the text to examine the probability of it being sunspots/auroras.

The target period of this study is from 581 – 960 CE, including the period of *Suí* (581–618 CE), *Táng* (618–907 CE), and the Five Dynasties and Ten Kingdoms (*Wǔdàishíguó*)[2] (907–960 CE). The *Suí* and *Táng* dynasties (581–970 CE) placed their capitals at present day *Xīān*, though their contemporary names were *Dàxīngchéng* (大興城) and *Chángān*, respectively, except for the era of *Wǔzhōu* (武周, 691–704 CE), who placed her capital at *Shéndōu* (神都), i.e. present day *Luòyáng*. This period includes the 774/75 CE event (Miyake et al. 2013), and the grand Minimum Candidate circa 685 CE (Eddy 1977b; Usoskin et al. 2007). After the fall of the *Táng* dynasty, *Hòuliáng* (後梁, 907–923 CE) placed her capital at *Luòyáng* and following dynasties (923-960 CE) placed their capitals at *Biàn* (汴), i.e. present day *Kāifēng*.

   We made complete surveys of sunspot and aurora candidates in five official histories which cover the periods described above:

    1. *Suíshū*, compiled by *Wèi Zhēng* and *Chángsūn Wújì* in 656 CE (*Suí* Dynasty: 581–618

---

[1] The term of "*wǔxíng* (五行)" is translated on one hand to "Five Elements" by Bielenstein (1950; 1984), Hayakawa et al. (2015), and Kawamura et al. (2016), and on the other hand translated to "Five Phases" by Pankenier (2013). In our study, we use translation of "Five Elements" to keep consistency with our previous papers (Hayakawa et al. 2015; Kawamura et al. 2016).
[2] This is a historical term which indicates a series of interval periods between the *Táng* and *Sòng* dynasties.

CE[3])

2. *Jiùtángshū*, compiled by *Liú Xù*, etc. in 945 CE (*Táng* Dynasty: 618–907 CE)

3. *Xīntángshū*, compiled by *Ōuyáng Xiū*, etc. in 1060 CE (*Táng* Dynasty: 618–907 CE)

4. *Jiùwǔdàishǐ*, compiled by *Xuē Jūzhèng*, etc. in 974 CE (the Five Dynasties and Ten Kingdoms: 907–960 CE)

5. *Xīnwǔdàishǐ*, compiled by *Ōuyáng Xiū* in 1053 CE (the Five Dynasties and Ten Kingdoms: 907–960 CE)

It should be noted that the official histories of these dynasties had been compiled after their ends, based on many kinds of their contemporary official documents. Especially of note is the fact that the *Táng* dynasty established *Shǐguǎn* (史館, the office of history), such that official histories were compiled under the leadership of following dynasties (*Jiùtángshū*, Staffs II: 1853; *Xīntángshū*, Staffs II: 1215). This is partly why one official history sometimes has records that are not written in another official history, as we can see later in the tables of sunspots and auroras.

The astronomical records in question in these official histories are compiled into and are available in the imperial chronicles (本紀), the Treatises of Astronomy (天文志), and the Treatises of the Five Elements (五行志).

Keimatsu (1970–1976) claimed that records in the Chinese official histories can be regarded more objectively than those in many other historical sources, because trained experts made regular observations at specified locations and recorded the celestial phenomena alongside dates and often with detailed notes on motions, shapes, and colors. However, it should be noted that the Chinese astronomical observations in pre-telescopic age

---

[3] This period is that of the imperial chronicle (*běnjì*) of *Suíshū*. However, the miscellaneous treatises (*zhì*) include records of previous dynasties from *Liáng* (since 502 CE) according to orders of Emperor *Tàizōng* in the *Tàng* dynasty.

were also made for the purpose of "astro-omenology" i.e., a kind of astrological fortune-telling and as "diagnostics" for policy makers (*Xīntángshū*, Five Elements I: 872; *Xīntángshū*, Staffs II: 1215; Pankenier 2013), and hence may have suffered from political influences. Several contemporary manuals of astro-omenology such as *Kāiyuán Zhànjīng* (開元占経)[4] by *Qútán Xīdá* (瞿曇悉達, *Gautama Siddha), an Indian chief astronomer of the national observatory of *Tàng*, are still available (Sasaki 2013a; 2013b). Astronomical records with texts for astro-omenology are for example given as follows:

(EX1) *Xīntángshū*, Five Elements I: p893

**Original Text:** 景龍二年七月癸巳，赤氣際天，光燭地，三日乃止。赤氣，血祥也。

**Translation:** On 24 Jul. 708, a red vapor reached the sky, illumined the earth like a candle fire. After three days it disappeared. A red vapor is an omen of blood.

In order to deliver celestial messages to the emperors, observatories were known to have been placed in the imperial palace or nearby (*Suíshū*, Astronomy I: 505, 533; *Jiùtángshū*, Astronomy II: 1335-36) in the eras of the *Suí* and *Tàng* dynasties, and most likely also in the era of the Five Dynasties and Ten Kingdoms. Owing to their political importance, astronomical observations were made even during wars. For example, during the *Ān Lùshān* Rebellion (安史之亂) from 755–763 CE, which was the most momentous civil war in the reign of *Tàng* dynasty, considerable records of observations were made.

Due to these astro-omenological aspects, we should note that astronomical records in official histories are sometimes interpreted as baleful signs or portents and gathered to form

---

[4] For bibliographical studies of this literature and its critical edition, see Sasaki (2013a; 2013b)

indirect criticisms against the emperors or governments. Bielenstein (1950; 1984) gathered "unnatural" phenomena recorded as portents in Western and Eastern *Hàn* dynasties to examine their distributions to find different frequencies during the reigns of various emperors. Although we have evidence of the actual observation of these portents supported by several examples of simultaneous observations of auroras such as in record of 937/02/14 explained later. Similar examples of simultaneous observations of auroras are regarded as the most probable aurora candidates (see also, Willis et al. 1999; Hayakawa et al. 2016a; Hayakawa et al. 2016b; Kawamura et al. 2016). Nevertheless we need to be aware of the selection bias to make indirect criticisms to contemporary emperors[5].

## 2.2. Search Method

We searched for passages in these official histories that included the keywords associated with sunspots and auroras using the digital search engine, Scripta Sinica (http://hanchi.ihp.cinica.edu.tw), provided by Academia Sinica in Taiwan (http://www.sinica.edu.tw). The target keywords are "black spots (黑子)" and "black vapors (黑氣)" in the sun for sunspots and "vapor (氣)", "cloud (雲)", "light (光)"[6] with color for auroras. Once the sentences that include the keywords were selected, we read corresponding sections of the original text to check if they actually refer to sunspots or auroras to remove unsuitable ones (e.g. those observed during day-time and so on). In the case where the record has the date of observation, we also calculated the moon phase to determine the sky conditions.

---

[5] In the same time, Yang et al. (1998) claim that there is no correlation between astronomical records and political events examining distributions of meteor shower records in Korean official histories. His result and Bielenstein's studies show us that it is a great topic to reconsider the correlation between criticisms on governments and astronomical records in official histories.

[6] Compared with aurora records, sunspot records concentrate in the Astronomical Treatises and we searched for relevant words throughout those chapters.

**2.2.1 Sunspot Records**

In the Astronomical Treatises of these official histories, sunspots are categorized in the subsection of unusual phenomena in the sun[7]. Typically, the sunspots are recorded as "black spot/vapor (黑子/黑氣) in the sun". Sometimes these records are accompanied by information about the number, shape, and size. For example:

(EX2) *Xīntángshū*, Astronomy II: p834.

**Original Text:** 開成⋯二年十一月辛巳，日中有黑子，大如雞卵，日赤如赭，晝昏至于癸未。

**Translation:** On 22 Dec. 837, a black spot was in the sun, as large as a chicken's egg, the sun was as red as red soil, and daylight was dark until 24 Dec.

The sizes are described in comparison with a tangible object, such as "a chicken's eggs" or "glasses". We do not know their relative nor absolute size indicated by these expressions.

It is possible that the sunspot observations may have been recorded without using the keywords we have searched. For example, Saito & Ozawa (1992) regarded "the sun was weak and without light" as the record of a sunspot. Such sunspot candidates are not included in our list.

**2.2.2 Auroral Records**

Although the pre-modern Chinese did not know the physical nature of aurora, there are considerable numbers of records that can be considered as aurora observations. For example,

---

[7] Records of sunspots can be found independently in the imperial chronicles during the same period.

Keimatsu (1969a; 1969b), Yau et al. (1995), and Saito and Ozawa (1992) claimed that the word "vapor" is likely to indicate auroras. Keimatsu (1970–1976) listed all the luminous phenomena seen at night and also discussed whether they correspond to auroras or not. However, Yau et al. (1995) indicated that Keimatsu's work mistakenly included comets or shooting stars to focus their target on terms such as "red vapor". Therefore, we assume that the records of luminous phenomena observed at night are potentially those of auroras and surveyed the words that refer to those phenomena: vapor, light, and cloud. In the Imperial Chronicles, the Treatise of Astronomy, and the Treatise of the Five Elements of the official histories in the abovementioned dynasties. From the list of potential aurora candidates, we manually removed the following two types. The first are those without dates. Most of these are found in the treatises explaining how astro-omenology is performed, which includes conversations between the emperor and his officials. Therefore, they are unlikely to be direct records of observations. The second are those seen during the daytime. Some of the daytime phenomena are presumably halos around the sun.

Records of aurora candidates often include information about their color, motion, and direction, the length, shape, and number of their stripes, and sometimes the location of the observation when it was not made in the capital city, i.e. *Rùnzhōu* or *Yángzhōu*. Examples of aurora records are given as follow.

(EX3) *Xīntángshū*, Astronomy II: p836

**Original Text:** 大曆十年…十二月丙子，月出東方，上有白氣十餘道，如匹練，貫五車及畢、觜觽、參、東井、輿鬼、柳、軒轅，中夜散去。

**Translation:** On 12 Jan. 776, above the moon in the eastern sky, there were more than ten bands of white vapors like a piece of silk, penetrating Tauras, Orion, Cancer, Hydra, and Leo, and they disappeared in the midst of night.

(EX4) *Xīntángshū*, Five Elements I: p894

**Original Text:** 寶應元年八月庚午夜，有赤光亙天，貫紫微，漸移東北，彌漫半天。 **Translation:** On 1 May 762, at night there was a red light filling up the sky, penetrating near the north celestial pole, gradually moved toward the north-eastern direction to spread to the half of the sky.

Their color is described as white, red, literally bluish[8], yellow, or a mixture of these colors. The colors are likely to be associated with what traditional Chinese called *Wǔxíng* or the Five Elements (五行), the theory in which the world consisted of Five Elements, namely metal, fire, wood, soil, and water, whose metaphor was shown in colors of white, red, literally bluish, yellow and black, respectively[9]. Thus appearance of vapors, lights, or clouds in these colors were considered as omens or metaphors of corresponding phase; metal, fire, wood, soil, or water. Their motions and directions are usually given by the eight points of the compass. Some records include information about constellations, planets, or the moon accompanying the auroras; EX3 is one of typical examples. Their lengths are given in units of *chǐ* or *zhàng*. In these eras, scales were different from dynasty to dynasty, although 1 *zhàng* was constantly 10 *chǐ* throughout these dynasties. 1 *chǐ* was equal to 29.51 cm in the first half of *Suí* (581–602 CE), 23.55 cm in the latter half of the same dynasty (603–618 CE), and 31.10 cm throughout the periods of *Táng* and the Five dynasties and Ten kingdoms (Tonami et al. 2006). At this moment, we do not know how the lengths expressed using these units correspond to what was actually seen in the sky. The shape and number of bands are also described only in some of the recorded events. The shapes are expressed in a figurative way, such as "like a rainbow" or "like a silk textile."

---

[8] The color described in characters of "cāng (蒼)" or "qīng (青)" in Chinese historical documents are sometimes used to represent "green" color (Pankenier 2013). However, we conventionally translated these characters as "literally bluish" throughout our article, to distinguish them from "literally greenish (綠)" that are frequently found in *Qīngshǐgǎo* (清史稿) (c.f., Kawamura et al. 2016).

[9] Therefore, red vapor was considered to be a metaphor of fire, for example. As for the detailed discussions on the metaphor of colors worldwide, see Lakoff & Johnson (1980; 2003).

## 3. Results and Discussion

### 3.1 Overall result

The lists of the aurora candidates are shown in Tables 1 (*Suí*), 2 (*Táng*), and 3 (Five Dynasties &Ten Kingdoms), and the list of sunspot candidates are shown in Table 4, 5, and 6. In total, we found 16 sunspot candidates (black spot/vapor/light/etc. in the Sun) and 45 aurora candidates (vapor/cloud/light during the night). The list is also available at our website: http://www.kwasan.kyoto-u.ac.jp/~palaeo/.

Figure 1 shows the annual number of records of candidates of sunspots and auroras. It is difficult to see any signature of the 11-year solar cycle from these records, but we can observe some long-term modulations. Overall, the period of 640–710 CE (the Minimum Candidate (Eddy 1977b; Usoskin et al. 2007)) appears to be less active than the remainder. In particular, the aurora candidate records cluster around the periods of 760–780 CE and 820–840 CE. In 820-840 CE, some sunspots are observed as well. However, no sunspot candidate was found during this period.

Within 45 aurora candidates, the records of 567 (month/day: unknown), 786/12 (day: unknown in Chinese record, 19 in Link catalugue), 827 (month/day: unknown in Link Catalugue), 937/02/14 were recorded simultaneously in Chinese and in Normandy in Europe, especially, one on 937/02/14 is the same in day-level. These records were with higher probablities as aurora candidate.

(EX5) 937/02/14: *Jiùwǔdàishǐ*, *Jìnshū*, *Gāozǔ*: p994[10]

**Original Text:** 天福二年春正月乙卯⋯是夜，有赤白氣相間，如耕墾竹林之狀，自亥至丑，生北濁，過中天，明滅不定，徧二十八宿，徹曙方散。

**Translation:** On 14 Feb. 937, at night red and white vapors appeared alternately, like a cultivated and exploited bamboo forest, from 23:00 to 3:00, muddily from north to the middle in the sky, flickering unstably went around the 28 lunar mansions[11] and disappeared at the dusk.

(EX6) 937/02/14, *Historia Ecclesiastica*, III: p146[12].

**Original Text:** Secundo post hæc anno, XVI kal. martii, circa gallorum cantum usque illucescente die, sanguineæ acies per totam cœli faciem apparuerunt.

**Translation:** Two years later from this year (935), 14th February, around the cockcrow, continuously illuminated by day, bloody light appeared through all of sky.

H15 reported 193 aurora candidates and 38 sunspot candidates in the official history of the *Sòng* dynasty (960–1279 CE); 0.6 event a year. In comparison, the number of candidates in found in 581–959 CE is smaller; 0.1 event a year. The *Sòng* era overlaps with the so-called Medieval Warm Period (MWP: 10–13 century) when the solar activity was stronger. Besides, the geomagnetic latitude of China was likely to be higher in the period of *Sòng* than in the periods of *Suí*, *Táng*, and the Five Dynasties and Ten Kingdoms (Butler 1992; Kataoka et al. 2016, submitted). These solar and terrestrial causes may be reflected in the larger number of aurora/sunspot candidates during the period of *Sòng,* though the possibility of political and human causes cannot be ruled out.

---

[10] Yau et al. (1995) and Xu et al. (2000) mistakenly relate the reference of this record with ch. 76 *Xin wudai shi* (*Xīnwǔdàishǐ*) but this is clearly their mistake with *Jiùwǔdàishǐ*. *Xīnwǔdàishǐ* has only 74 volumes after all.

[11] In traditional Chinese astronomy, the heaven was separated into four directions and every direction has seven lunar mansions in it. Thus, this statement means this "vapor" floated around the whole sky.

[12] Orderici Vitalis, *Historia Ecclesiastica*, Parisiis, 1845.

**3.2 Color of auroras**

Among the 45 aurora candidates, 17 are white, 21 are red, and 8 are other colors, such as blue, blue-red, or red-white. It should be noted here that the latitude of the locations of the observations, which in most cases are at the contemporary capital, is considered to be not low enough for low latitude.. Fig.2 shows that the temporal evolution of the geomagnetic latitude of *Kāifēng* and *Xīān*. Using global geomagnetic field model CALS3k.4b (Korte & Constable, 2011), geomagnetic latitude of observation cite in China in this term is about 34-38 degrees.

Generally, the color of auroras seen in such low latitude is red. The color of the aurora reflects the state of atoms and molecules in the upper atmosphere, and energy of precipitating electrons. For forbidden lines of atomic oxygen, green aurora at 557.7 nm, which is the most common auroras, are seen in the mid-altitude range (100–150 km), and red one at 630.0 nm is seen at high altitudes (>200 km). Blue and violet auroras at 427.8 nm and 391.4 nm associated with emission of $N_2^+$ coincide with the green aurora. In very intense auroras, blue, violet, or pink colors at low altitudes (80–100 km) are sometimes seen which come from the molecular nitrogen lines ($N_2$) in the blue and red parts of the emission spectrum. These lines are mostly excited by energetic electrons injected from the magnetosphere (Chamberlain 1961). If seen from a latitude lower than that of the auroral oval, only the upper part of the aurora, where the red (630.0 nm) line dominates, is visible. It is also known that, during large geomagnetic storms, so-called stable auroral red arcs (SAR arcs) become visible in the mid latitude range (Rees and Roble 1975, Kozyra et al. 1997).

It was already pointed out in H15 that the pre-telescopic Chinese were likely to have expressed the color according to the *wǔxíng* (五行) or the Five Elements, in which the term "green" was not used. We have two possibilities to explain this fact. The first one is that green auroras were described as those in "literally bluish (蒼/青)" color. It is not rare for green objects to be described as "literally bluish" in pre-modern China (see, Pankenier 2013). However, this theory cannot explain everything as there are only three "literally bluish" aurora candidates within official histories in these period. The second one is that the term "white" alongside auroral candidates in the Chinese records correspond to the green color of the auroras from the oxygen 557.7 nm line. It is also possible that very faint green auroras were actually seen as "white" auroras. In the case of very faint auroras, the cone cells in human eyes, which are responsible for color vision, are less responsible, and therefore the rod cells, which are used in peripheral vision, dominate (Purkinje effect, Purkinje 1825). A rod cell is sensitive enough to respond to a single photon of light, that is, at night-time. A rod cell is the most sensitive to wavelengths of light around 498 nm, therefore the faint 577.7 nm emission is seen as white. These color-shift effect is known at naked-eye observation of stars (Cheng 2009) and therefore the same faint aurora can be seen white or green. Another explanation would be possible in terms of mixing colors.   In many cases, the green auroras (557.7 nm) coincide with a number of emission lines from ultraviolet to infrared, in particular, the blue and violet auroras at 427.8 nm and 391.4 nm ($N_2^+$ first negative) and red ones around 670 nm ($N_2$ first positive) at similar altitudes.
If the red aurora (630.0 nm) occupied the background, one would see the aurora as "white."

In order for the green auroras to be seen in the Chinese capitals, however, the aurora oval would have had to extend to relatively low latitudes (30–35N). It is rare but possible that during extremely intense geomagnetic storms, such as the Carrington event of 1859 and a recent event of 1989 (Kimball 1960; Allen et al. 1989; Hayakawa et al. 2016b), the aurora oval may expand to such low latitudes. Considering that the geomagnetic latitude of China in the periods covered in this paper was probably higher than that of present China, it is not unreasonable to think that at least some of the white aurora candidates were indeed the green (white) auroras.

This second possibility is well supported by historical records for auroras in great magnetic storms. In the Carrington event in 1859, we have a dozen reports to describe green auroras as "white/whitish" (Loomis 1860a, b; 1861). In drawing of *Enkoan Zuikai Zue* (猿猴庵随観図会)[13] for aurora observation on 1770 Sep 17 in Japan, it is described as "white stripes in red vapor" as well as the records of the simultaneous observation in Fushimi (Nakazawa et al. 2004; Iwahashi 2016, submitted; Tamazawa et al. 2016, submitted). This description is very similar to that of record in 937 (EX5) that is confirmed as an aurora thanks to simultaneous observation in Normandy (EX6).

However, that the number of the white aurora candidates (21) is comparable to that of the red candidates (18) is still puzzling, as the green (white) auroras at such low latitude is much rarer than the usual red auroras at low latitude. It is hence likely that the other phenomena are also included in our list, particularly in the "white" auroral candidates. One possible non-auroral origin is the atmospheric optics events. Similar to our previous study (H15), here we present the result of the lunar age analysis. The atmospheric optics of the moon is a phenomenon in which ice crystals shaped as hexagonal prisms are suspended in the air and reflect or refract the moon light. This phenomenon is commonly known by its partial structures, such as sundogs, parhelic circles, and a 22-degree halo. Our hypothesis is that records would be distributed around the period of the full moon if a significant amount of atmospheric optics records contaminated the observational records. The result is shown in

---

[13] National Diet Library of Japan, MS.特 7-59, ff. 6b-7a.

Fig. 1, from which we could not decode any significant trend with the 36 records against the lunar age.

The difference of timing of observation or expression of phenomena (氣, 光, 雲) may have information to judge whether they are aurora or not. Fig. 4 and 5 show short peaks of observational records in July and October exist in Táng era. It is difficult to speak about a July peak from the difference of recorded color or expression. Simply thinking, the longer the night time is, the more chance of observation is. Therefore a part of records in July may include non-aurora records.

### 3.3 Moon Phase Calculation

The astronomical records in official histories in the pre-telescopic age include candidates of atmospheric optics, such as "lunar halo" or "paraselene" (or "moon dog"), as well as aurora candidates. The moon phase is a trait of the sky that we can calculate accurately for historic dates. To do this, we referred to the 6000-year catalog of moon phases with Julian dates found at the NASA Eclipse website (http://eclipse.gsfc.nasa.gov/phase/phasecat.html). This catalog is based on an algorithm developed by Meeus (1998).

Fig.3 are record histograms concerning the normalized lunar age (0.0, 0.5 and 1.0 means the new moon, the full moon, and the next new moon). No significant tendency is found in these histograms. This tendency (no significant tendency) is the same as Chinese official histories of 10th -13th century (Hayakawa et al. 2015) or 14th-16th century (Hayakawa et al. 2016e, submitted).

### 3.5 $^{14}$C event in CE 774/775

As is already mentioned in section 1, evidence of sharp increase in cosmic ray flux during CE 774-775 are found in the tree rings (Miyake et al. 2012; Usoskin et al. 2013; Jull et al. 2014;

Güttler et al., 2013) and the corals of the Chinese Sea (Liu et al., 2014). Miyake et al. (2015) also reported a clear increase of $^{10}$Be concentration in CE 775 in the Dome Fuji ice core. Motivated by this event, Stephenson (2015, hereafter S15) reviewed Chinese official histories from CE767 to CE776, and Chapman et al. (2015, hereafter C15) reviewed the East Asian records of auroras from CE757 to the end of the 770s reported in the contemporary documents[14]. While we include all the entries in S15 except for one for black vapor (黒氣) on 21 Jan 768 that is not within our target keyword for aurora candidates, we have several difference with lists shown by C15. Since the period is included in our survey, here we briefly present a cross-check of their result and ours.

### 3.4.1 "Misinterpreted" auroras in C15

Seven "misinterpreted" auroras, i.e. the records that were claimed to be aurora candidates by previous authors but actually unlikely to be auroras, are listed in the section 2 of C15. Of seven records of "misinterpreted" listed in C15, two records, 763 Nov and 767 Jul/Aug (2.2 and 2.4 in C15), are not from Chinese official histories, therefore these records are not included in our paper. Also, the record of 774 Oct 13 (2.6 in C15) is not listed in this paper because it does not include the terms we used for the search of aurora candidates. C15 pointed out that this event was listed in the astronomical treatise probably because these records mention the Moon, Mars, or Venus.

---

[14] Although Neuhäuser & Neuhäuser (2015, hereafter NN15) also reviewed contemporary Chinese records in wider range, we do not compare our catalogue with this study as their catalogue is not based on the original text i.e. only a secondary catalogue based on previous catalogues. NN15 explicitly state that they "reviewed scholarly translation of Arabic, Syriac, and **Chinese**" in the section of "3.1. aurora criteria and sources". Seeing their reference list, we can find Keimatsu (1970-76), Yau et al. (1988; 1995), Xu et al. (2000), and Liu (1975; 2007) as relevant references. Within these references, the original historical sources are only given as "Liu (1975; 2007)" i.e. *Jiùtángshū* but we could not find *Xīntángshū* (possibly written as "Ouyang 1975" in NN15's style as in C15) in their references. Additionally, we need to note that *Jiùtángshū* is completed by *Liú Xù* (劉昫) but started *Zhāng Zhāoyuǎn* (張昭遠) with aids of their colleagues in the department of history (史館) (*Jiùwǔdàishǐ, Jìnshū* V: p1046). Hence their reference should be corrected to Liu et al. (1975; 2007). On the other hand, Keimatsu (1970-76), Yau et al. (1988; 1995), and Xu et al. (2000) are catalogues of records of aurora-like phenomena and sunspots with English translations as we explained above. Considering these editions (Liu 1975; Liu 2007) are written only in Chinese without European translations and above-mentioned statement in "3.1. aurora criteria and sources" by NN15, it is quite sure that they did not consult these Chinese original texts but catalogues with English translations. Therefore, NN15 is no more than a secondary catalogue recompiling previous catalogues and it is of no use to compare our results with NN15. Their attitude to ignore original text in oriental languages is not only for Chinese historical documents in NN15. NN15 also has problems on interpretations for Syriac chronicles as well that are hence corrected by Hayakawa et al. (2016c).

The remaining four records are listed in this paper as well; 761 Dec 13, 767 Aug 25, 773 Aug 9, and 776 Jan 11/12. C15 judged those on 767 Aug 25 and the 773 Aug 9 (2.3 and 2.5 in C15) "misinterpreted" because these events were recorded as white vapor (白氣) at twilight (in original text, "酉時"), whereas those on 761 Dec 13 and the 776 Jan 11/12 (2.1 and 2.7 in C15) "misinterpreted" because these events were likely atmospheric optics, i.e. phenomena associated with Moon (Moon Halo). However, intense auroras may be visible with naked eyes even during the twilight or in the presence of the moonlight.

First of all, as we saw above, the chronicler Orderic Vitalis reported an aurora-like phenomena appeared around the twilight time (gallorum cantum) (EX6, *Historia Ecclesiastica*, III: p146) simultaneously with another Chinese report (EX5). We have another examples for simultaneous aurora observation in 1363 as well. In this time, we have simultaneous aurora observations on July 30: "at sunset (日暮)" in China and "at the beginning of night (入夜)" in Japan (Willis et al. 1999; Hayakawa et al. 2016a).

Secondly, we can find numerous reports on the aurora seen during the twilight associated with the famous "Carrington event" in 1859 (Shea and Smart 2006):

> "While the evening twilight was yet so strong as to make the phenomenon scarcely discernible, a rosy hue was seen spreading over a space reaching from the northeastern horizon to the north star and thence to my zenith, of uniform breadth throughout, and bounded south by a line through Alpha Lyrae, passing vertically down to the east.

(Observations of Prof. ALEXANDER C. TWINING on the Aurora of Aug 28th, 1859, made at West Point, New York)",

"Aug. 28th, the aurora was visible in the evening twilight especially to N. and N.E. (Observations at Gettysburg, Pennsylvania, (lat. 39° 49' N, long. 77°15' W), by Rev. M. JACOBS)",

and

"The aurora continued till daylight, when it gradually faded away… This was most vivid from 1h15m to 1h 45m, but was continued till almost daylight. (5. Observations at Grafton, Canada West (lat. 44°3′ N. long. 78°5′ W), by JAMES HUBBERT.)".

Also, there are numerous records of the red aurora (and some records reported white stripes in it) observed on 1770 Sep 17 in various area across Japan, and in some records indicate the "red vapor" appeared during the twilight time (酉時) (Nakazawa et al. 2004). C15 is clearly ignoring these concrete reports for aurora observations in twilight time.

Furthermore, although it is relatively more difficult to see an aurora around full moon, the possibility depends on, at least, brightness of aurora, and angular distance between Moon and sky position. It is also confirmed in modern scientific literature as well. Barnard (1910), for example, describes aurora seen at night with full moon as follows: "September 4 (1908). 7h 30m: Bright aurora with nearly full moon. … 10h40m: The arch was very bright in spite of a bright moon." It is notable as well that several observers of aurora displays in 1859 compared the aurora light equals with that of "a full moon" and even some regard the aurora light was exceeding the full moon (Loomis 1860a,b; 1861) as is reported in the *Baltimore American and Commercial Advertiser*[15]. Hence, it is not too difficult to see auroras even at night with full moon and it is too speculative to relate aurora-like records at the night with lunar halo just because it was seen at night of full moon.

---

[15] "The Aurora Borealis", p2., column 2 in the paper of 1859 Sep 03.

In particular, the record of 776 Jan 11 or 12 includes information of the area in the sky; the white vapor was seen in the area whose elevation angle was up to 40°- 90° (the zenith) above the horizon, with its direction east-south-west. Therefore, we should note that this record cannot be explained by lunar halos (suggested by C15) neither with a radius of 22° nor 46° and have considerable possibility to be an aurora observation, although its lunar phase suggested that the moon was quite bright. Keimatsu (1973) listed this event as a "very probable" aurora and S15 identify this record as aurora observation, whereas Yau et al. (1995) did not listed it as an aurora.

We do not, however, claim that the above four records listed in our paper but labeled as "misinterpreted" by C15 were definitely auroras. It is also possible that these records were "misinterpreted" as claimed by C15. However, they are still included in our list because the aim of our paper is to survey the potential candidates of the aurora candidates from the well-defined source documents and the criteria and provide the result to the scientific community for further investigation.

### 3.4.2 likely true auroras in C15

C15 listed only three events as "likely true" auroras (section 3). All of the three records are included in this paper.

### 3.4.3 Questionable auroras in C15

C15 listed five events as "questionable" auroras (section 4). The event of 760 Jul or Aug (4.2 in C15) is listed in this paper.

In this paper, the terms we resurvey as aurora candidate are vapor (氣), light (光) and cloud (雲) that are confirmed to mean auroras with simultaneous observations (e.g. Willis et al. 1999; Kawamura et al. 2016; Hayakawa et al. 2016b). In several previous surveys for aurora records, other keywords are target of survey. Keimatsu (1960-1967) listed records with term such celestial dogs (天狗), rainbow clouds (五色雲), and so on, as aurora candidate (each number of record included these terms are smaller). Strictly speaking, Keimatsu (1960-1967) listed record of all the luminous phenomena in the sky as aurora candidates. On the other hand, Yau et al. (1996) did not list records in these terms. Records on 757 Feb 20 and 767 Oct 8 are described in the term white rainbow (白虹) and white mist (白霧). As Hayakawa et al. (2016a) discuss relation between the term "white rainbow (白虹)" and aurora in this era, we do not get in detail in this paper.

We listed the records on 770 Jun 20 and 770 Jul 20 (4.4 and 4.5 in C15) as two events, C15 suggested there is possibility that these two records indicate same date because the descriptions for these records are alike and thus have a possibility of misdating. Although it is possible, it is no more than a speculation in the viewpoint of historical studies, otherwise we get further source documents for these records.

### 3.4.4 Aurora candidates not mentioned in C15

From the list of aurora candidates in this paper, 763 Sep (day: unknown) event, are not listed in C15, and not listed in Keimatsu (1973) nor Yau et al. (1995). This events were red event, observed at night-time and northern direction, typical for very likely auroras.

### 3.5 Distribution of sunspot records

Within 19 sunspot records, we found most of them are found after 826 except for two in 567 and 577 and no records between 578 and 825. In astronomical terms, on one hand, this gap is partially explained with a Minimum candidate in 640–710 CE (Eddy 1977b; Usoskin et al. 2007). After the 820s, there was a high level of solar activity as seen in Fig 7. In this period, it is not only in China but also in Arabic countries and in Japan as well that naked-eye sunspots were observed (Wittmann 1977). On the aspect of the court development, on the other hand, we should pay attentions to several internal discords and civil wars. The most famous of them in this period is the *Ān Lùshān* Rebellion (安史之亂) during 755−763 CE. At this time, the rebels under *Ān Lùshān* once captured *Luòyáng* and *Chángān*, the capital cities of *Táng* dynasty (*Xīntángshū*, *Sùzōng* VI: 150-154; *Jiùtángshū*, *Xuánzōng* II: 230-34). Although *Táng* dynasty repressed this revolt to take back their capital cities, these capital cities were attacked and captured by invasion of *Tŭbō* (吐蕃) in 763 CE (*Jiùtángshū*, *Dàizōng*: 273; *Xīntángshū*, *Dàizōng*: 169). It is possible that some of archives of sunspot observations got lost, although it cannot explain all the absence of sunspot records up to 825 CE.

### 3.6 Comparison with reconstructed solar activity level

In order to examine the correspondence of the above results with solar activity, we compared the timing of the data with the proxy-based solar activity level. Figure 6 and 7 show the comparison with the solar activity level reconstructed with multiple cosmogenic nuclide records (Steinhilber et al. 2009). It is difficult to see the relation between long-term change of solar activity and records of aurora and sunspot candidates for 6-10th century. Few data of sunspot were recorded in the period of 640–710 CE, which is the Minimum Candidate (Eddy 1977b; Usoskin et al. 2007). Note that the absence of records does not mean less solar activity.

### 4. Conclusions

We have resurveyed sunspot and aurora-like records in the official histories of the dynasties during 581–960 CE. We found 16 sunspots and 45 auroral candidates during that period, with information of their size, color, etc. Our results are provided in this paper and on the web for use of the scientific community. No aurora candidate or sunspot records considered to have caused the 774/775 event (Miyake et al. 2012) are found in our survey. However, aurora candidate around 770s, not listed in other papers, should be considered because very huge low-latitude auroras are different from ordinary ones. We welcome and encourage the use of our data as well as the contributions that provide more data from diverse historical sources.

**Acknowledgements**

The authors wish to acknowledge Dr. Miyahara for useful comments. This research was supported by the following grants: the UCHUGAKU project of the Unit of Synergetic Studies for Space, Kyoto University; the research grant for Exploratory Research on Sustainable Humanosphere Science from the Research Institute for Sustainable Humanosphere (RISH), Kyoto University; the Center for the Promotion of Integrated Sciences (CPIS) of SOKENDAI; and Grants-in-Aid for Scientific Research (JP15H05816) of Japan Society for the Promotion of Science.

**Appendix: Information of critical editions for historical sources**

Here, we provide the information for critical editions for Chinese official histories with full respect for the manner of historians.

1. *Suíshū*: Wèi Zhēng & Chángsūn Wújì, *Suíshū* (隋書), Běijīng, I-VI, 1973.

2. *Jiùtángshū*: Liú Xù, et al., Jiùtángshū (舊唐書), Běijīng, I-XVI, 1975.

3. *Xīntángshū*: Ōuyáng Xiū, et al. *Xīntángshū* (新唐書), Běijīng, I-XX, 1975.

4. *Jiùwǔdàishǐ*: Xuē Jūzhèng, et al. *Jiùwǔdàishǐ* (舊五代史), Běijīng, I-VI, 1976.

5. ***Xīnwǔdàishǐ***: Ōuyáng Xiū, *Xīnwǔdàishǐ* (新五代史), Běijīng, I-III, 1974.

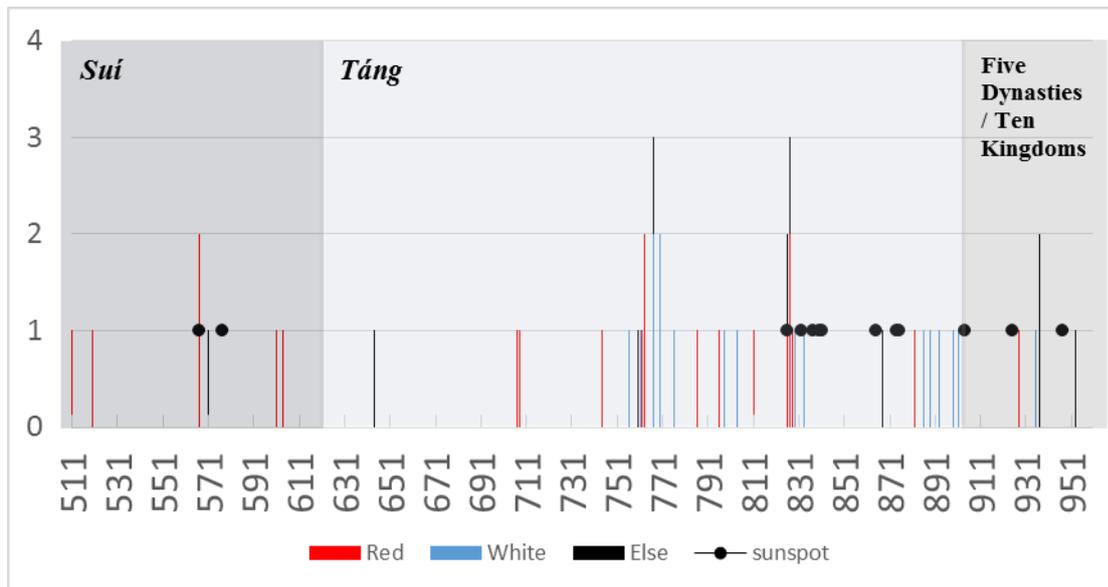

Fig. 1. Change in the number of white vapor, red vapor, and black spots during 517–975 CE. The bars in *blue* and *red* represent the number of candidates of white and red aurora, respectively. The *black dots* are for the total number of aurora candidates. The *crosses* show the number of sunspots. Note that records of sunspot and aurora candidate are from documents written in different era, therefore colors of background in this firure are used depending on era.

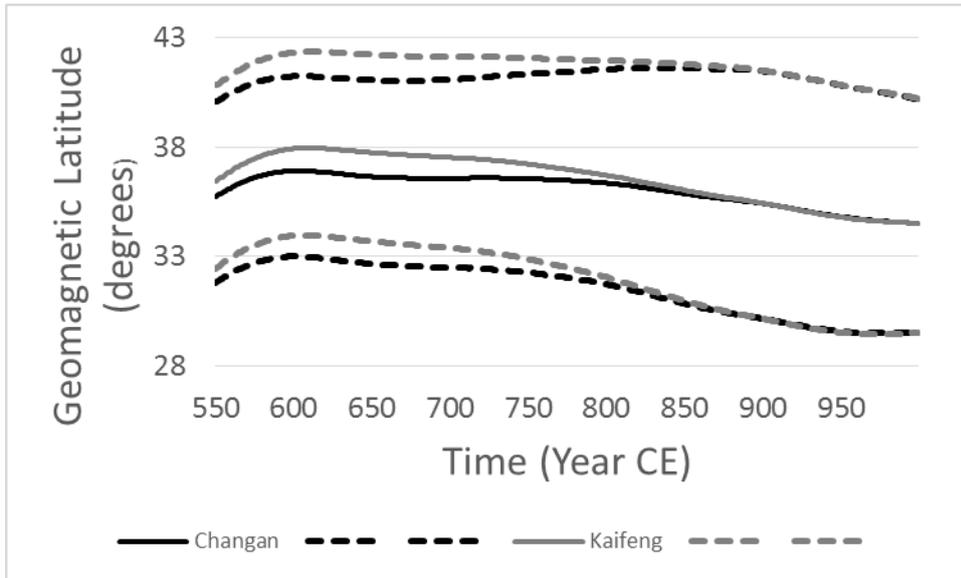

Fig 2. Time evolution of geomagnetic latitude at *Kāifēng* (black) and *Chángān* (glay) The error bars of one standard deviation as indicated by dotted lines, are evaluated by bootstrap method (Korte and Constable, 2011)

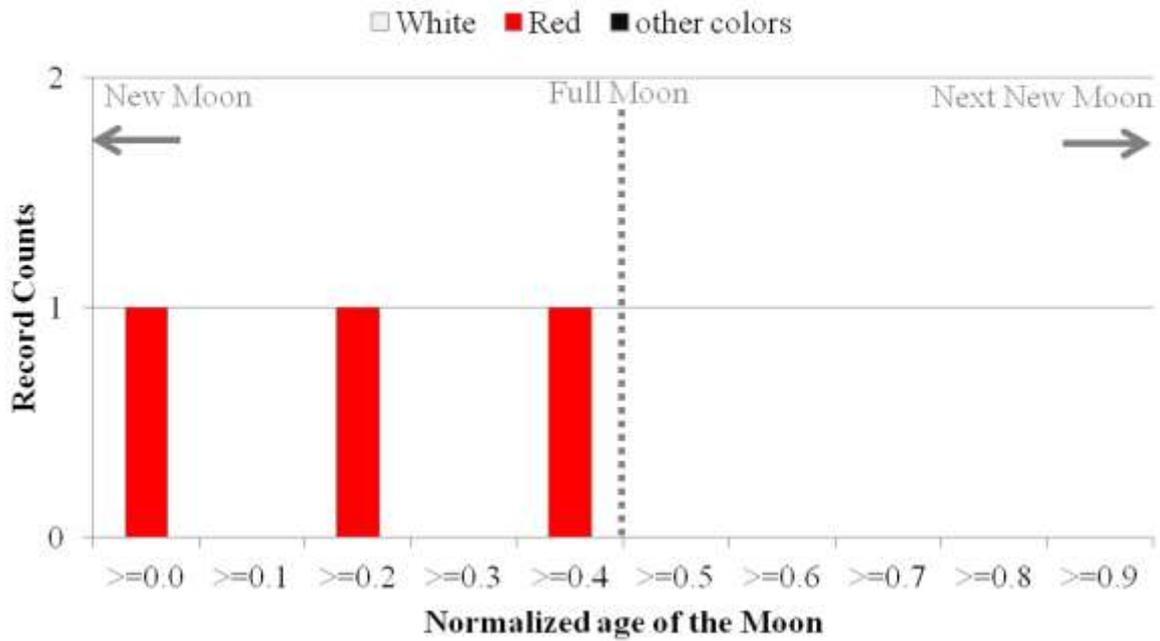

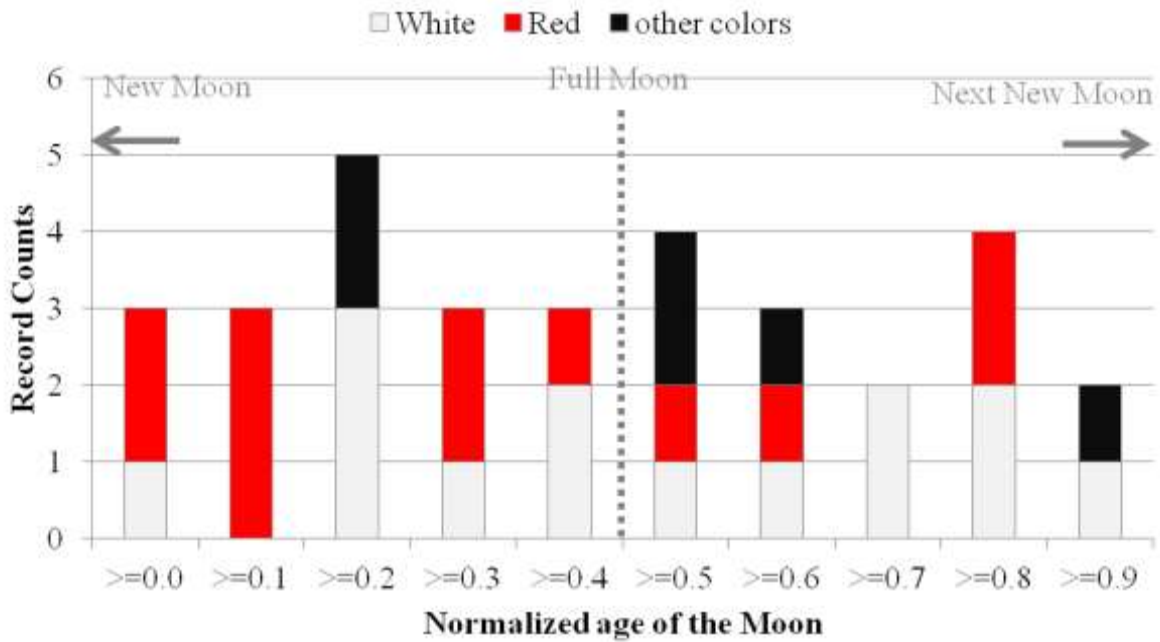

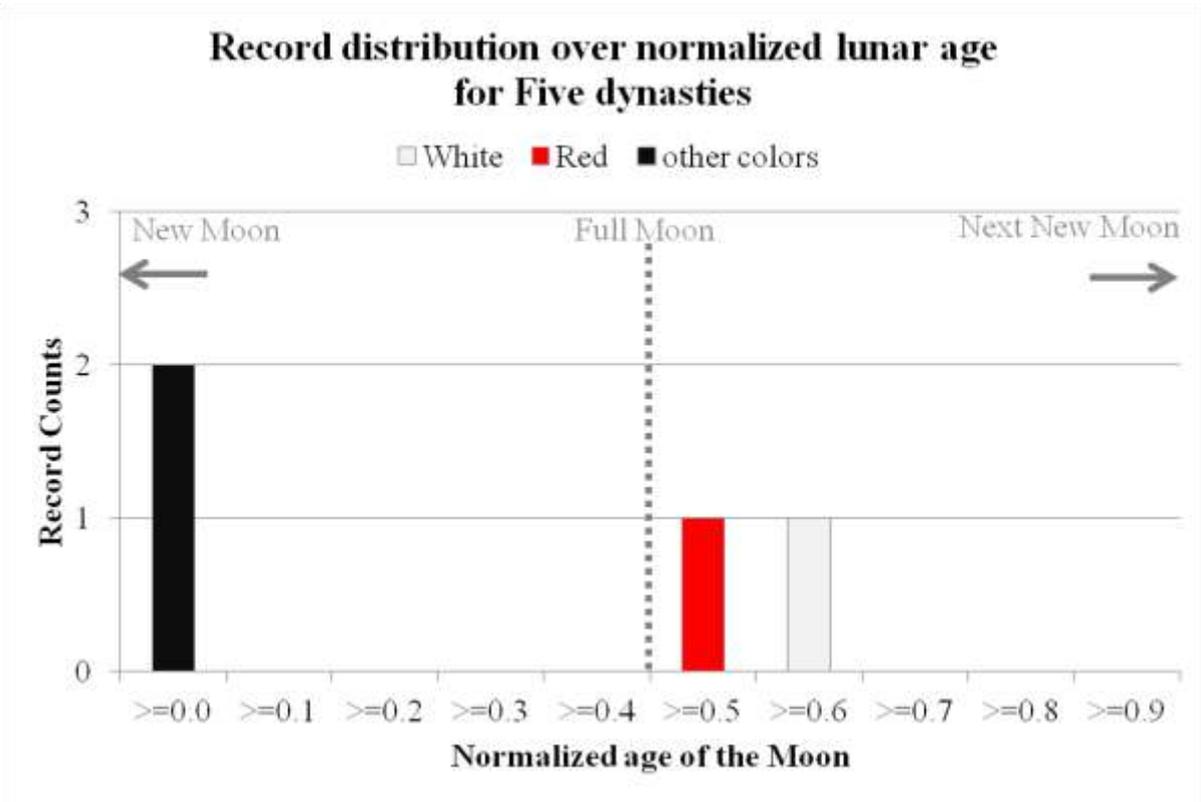

Fig. 3. Counts of aurora candidates against moon luminosity. This histogram is labeled with the reported color of vapor. The alphabetic labels represent the color of vapor as W (white), R (red), and all else (blue white)

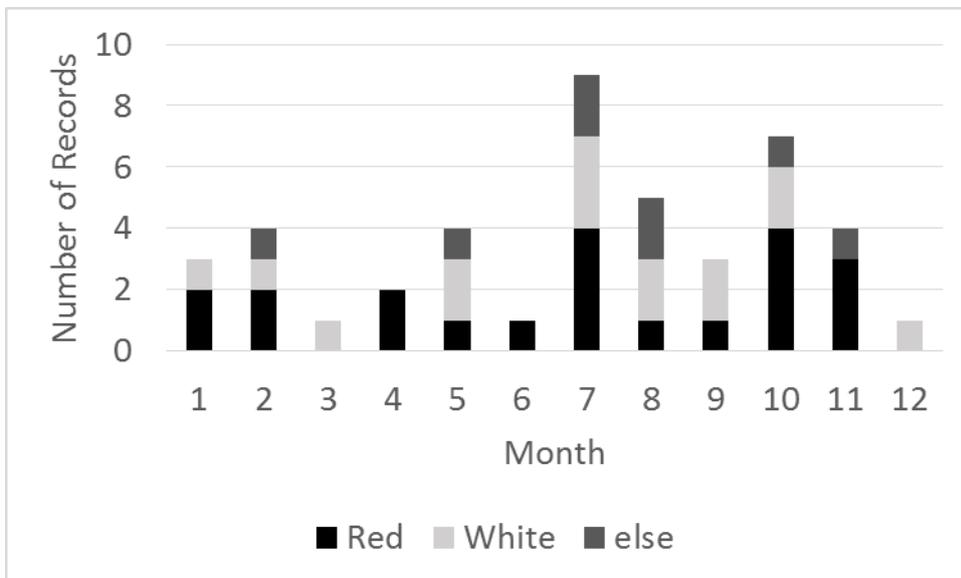

Fig 4 Conuts of auroral candidates against observed month in Tang term

. This histogram is laveled with the reported color.

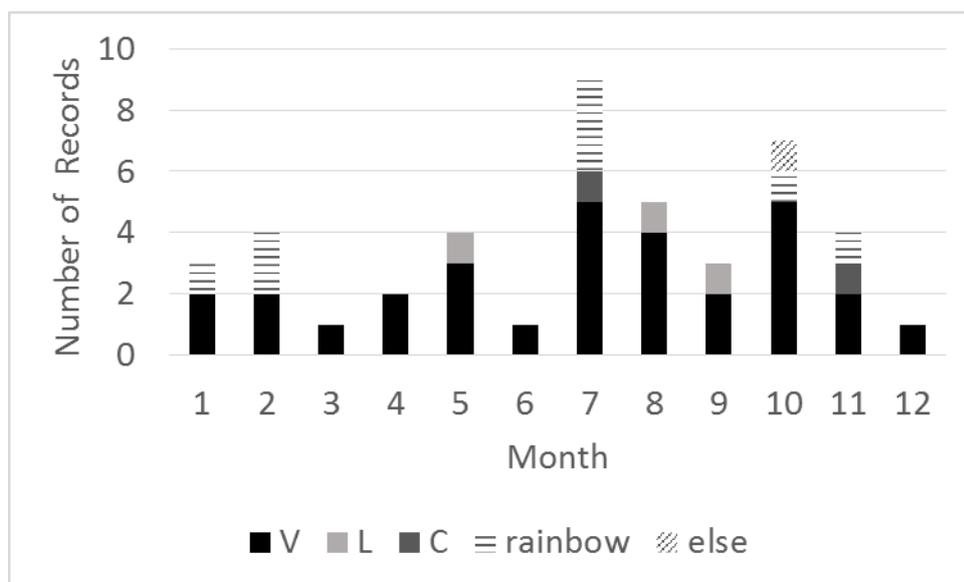

Fig 5 Conuts of auroral candidates against observed month in Tang term. This histogram islaveled with the reported expression; 氣(vaper; V), 光(light; L), 雲(cloud; C), 虹(rainbow), and so on .

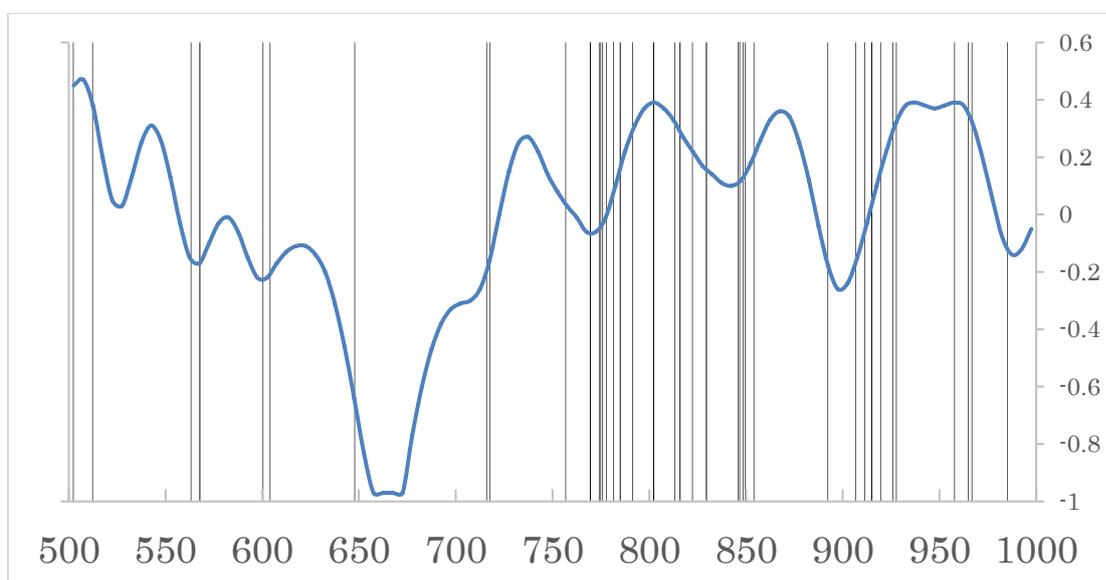

Fig 6 Records of aurora candidates (black lines) compared with solar activity from Steinhilber et al (2009)

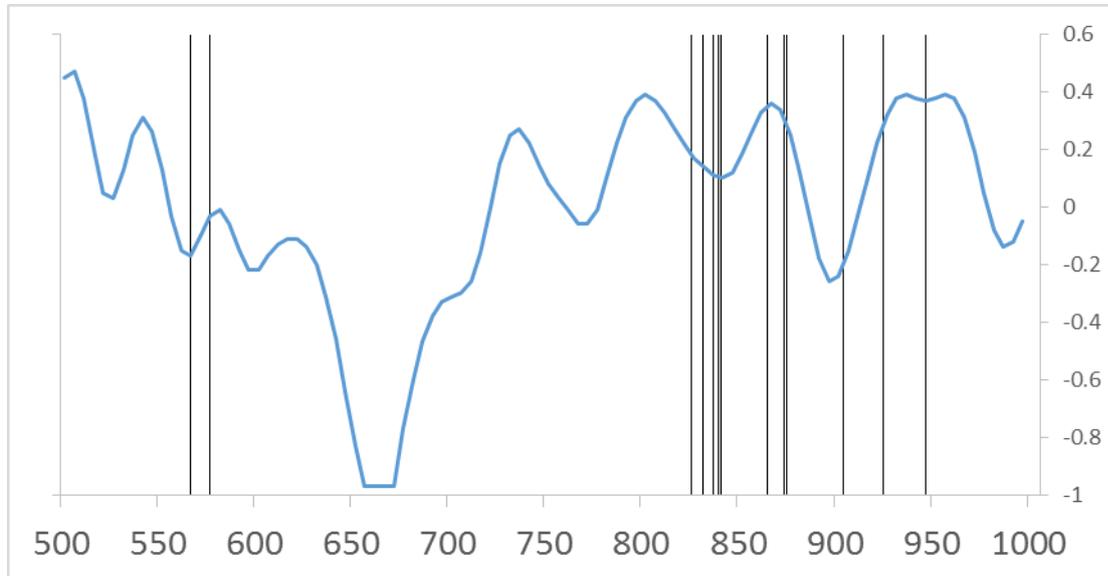

Fig 7 Records of sunspot (black lines) compared with solar activity from Steinhilber et al (2009)

Table 1. Auroral records of the *Suí* Dynasty

| Year | Month | Date | Color | Description | Direction | Length | counts | Place | Notes | Moon Phase |
|---|---|---|---|---|---|---|---|---|---|---|
| 511 | 10 | 10 | R | V | wn | | | | | 0.095225 |
| 520 | 10 | | R | V | wn | | | | | −1 |
| 567 | 5 | 31 | R | V | wn | | | | | 0.251852 |
| 567 | 11 | | R | V | wn | | | | | −1 |
| 601 | | | R | V | n | | | | | −1 |
| 604 | 1 | 21 | R | V | n | | | | | 0.490762 |

Table 2 Auroral records of the *Táng* Dynasty

| Year | Month | Date | Color | Description | Direction | Length | counts | Place | Notes | Moon Phase |
|---|---|---|---|---|---|---|---|---|---|---|
| 618~ | | | W | rainbow | | | | | | |
| 644 | 7 | | Bblack | V | | | | | | |
| 707 | 10 | 18 | R | V | | | | | | 0.608805 |
| 708 | 7 | 24 | R | V | | | | | | 0.089033 |
| 710 | 7 | 9 | | rainbow | | | | | | 0.281487 |
| 712 | 7 | | W | rainbow | | | | | | |
| 744 | 2 | 3 | R? | V | | | | | | 0.507632 |
| 756? | | | W | V | n | | | | | |
| 756 | 7 | 16 | W | C | wn | | | | | 0.489619 |
| 757 | 2 | 20 | W | rainbow | | | | | | 0.91884 |
| 762 | 5 | 1 | R | L | wn,en | | | | | 0.100398 |
| 763 | 8 | 24 | R | L | wn,e | | | | | 0.3729 |
| 763 | 9 | | R | L | | | | | | |
| 767 | 8 | 17 | BR | V | | | | | | 0.61636 |

| | | | | | | |
|---|---|---|---|---|---|---|
| 767 | 8 | 23 | R | V | | 0.818495 |
| 767 | 8 | 25 | W | V | | 0.885873 |
| 767 | 10 | 8 | W | ? | wn | 0.366718 |
| 767 | 10 | 25 | BR | V | | 0.938881 |
| 770 | 6 | 20 | W | V | wn | 0.759406 |
| 770 | 7 | 20 | W | V | wn | 0.783588 |
| 773 | 8 | 9 | ? | V | | 0.558167 |
| 776 | 1 | 12 | W | V | w | 0.560678 |
| 786 | 12 | | R | V | | |
| 796 | 10 | 20 | R | V | n | 0.504862 |
| 798 | | | W | V | | |
| 804 | 10 | 15 | W | V | e−w | 0.278408 |
| 811 | 3 | 31 | R | V | | 0.123109 |
| 819 | 1 | 6 | W | rainbow | e−w | 0.222969 |
| 826 | 1 | 21 | R | V | nw | 0.32588 |
| 826 | 2 | | R | V | n | |
| 827 | 5 | | R,W | V | n | |
| 827 | 7 | 22 | R | V | wn | 0.833763 |
| 827 | 9 | 8 | R | V | | 0.45842 |
| 828 | 5 | 17 | R | V | n | 0.007673 |
| 829 | 9 | | W | V | w | |
| 833 | 11 | 17 | W | V | w | 0.080332 |
| 844 | 2 | 17 | W | rainbow | w | 0.854538 |
| 860 | | | W | rainbow | w | |

| Year | Month | Date | Color | Description | Direction | Length | counts | Place | Notes | Moon Phase |
|---|---|---|---|---|---|---|---|---|---|---|
| 868 | 7 | 29 | W | rainbow | w | | | | | 0.218072 |
| 868 | 11 | 25 | ? | C | | | | | | 0.25701 |
| 882 | 7 | 24 | R | V | wn | | | | | 0.196711 |
| 886 | 5 | | W | V | es | | | | | |
| 886 | 10 | | W | rainbow | w | | | | | |
| 886 | 11 | | W | rainbow | w | | | | | |
| 889 | 4 | 25? | W | V | ws-en | | | | | |
| 893 | 11 | | W | V | | | | | | |
| 899 | 4 | 25 | W | V | ws-en | | | | | 0.404272 |
| 901 | 10 | 5 | W | V | w | | | | | 0.678495 |

Table 3. Auroral records of the Five Dynasties and Ten Kingdoms

| Year | Month | Date | Color | Description | Direction | Length | counts | Place | Notes | Moon Phase |
|---|---|---|---|---|---|---|---|---|---|---|
| 928 | 1 | 10 | R | V | ws | | | | | 0.50149 |
| 935 | 1 | 27 | W | V | ew | | | | | 0.677654 |
| 937 | 2 | 15 | R,W | V | * | | | | | 0.061861 |
| 937 | 2 | 14 | RW | | | | | | | 0.027875 |
| 953 | 5 | | Y | V | | | | | | |

Length: 1 *chǐ* = 29.51 in the first half of *Suí* (581–602 CE), 23.55 cm in the latter half of the same dynasty (603–618 CE), and 31.10 cm in *Táng* and five dynasties and ten kingdoms (Tonami et al. 2006).

Location of the observatories: *Dàxīngchéng* or *Chángān* (present *Xīān*) (34.16 N, 108.57 E), *Tōnghànzhèn* (unknown), *Luòyáng* (30.25 N, 120.17 E), *Rùnzhōu* (31.59 N, 119.35 E), *Yángzhōu* (34.25 N, 119.25 E), *Kāifēng* (34.80 N, 114.30 E).

Color of auroral candidates: *W* white, *R* red, *B* blue, *Y* yellow, *Bk* black, *R-W* red and white, *RW* red white, *BW* blue white, *Ra* rainbow; types of recorded phenomena: *V* vapor (氣), *C* cloud (雲), *L* light (光); direction: *e* east, *w* west, *s* south, *n* north, *m* middle, *en* northeast; these abbreviations can be combined: for example *wn-es* from northwest to southeast.

**Table 4** Sunspot records

| Year | Month | Date | description | Size | Counts | Place | Note |
|---|---|---|---|---|---|---|---|
| 567 | 12 | 10 | | | | | |
| 577 | 12 | | | | | | |
| 826 | 5 | 7 | BV | glass | | | |
| 826 | 5 | | BS | | | | |
| 832 | 4 | | B? | | | | |
| 832 | 4 | 21 | BS | | | | |
| 832 | 5 | 6 | BV | | | | |
| 837 | 12 | 22 | BS | chicken egg | | | |
| 840 | 12 | 28 | BV | | | | |
| 841 | 12 | 30 | BS | | | | |
| 865 | 2 | | BV | chicken egg | | | |
| 874 | | | BS | | | | |
| 875 | | | north dipper | | | | |
| 904 | 2 | 19 | north dipper | | | | |
| 925 | 12 | | | | | | |
| 947 | 11 | 26 | | | | | |

*BS* black spots (黒子), *BV* black vapor (黒氣), *WL* the sun was weak and without light (日淡無光), *shade* (景)